\documentclass[aps,prd,showpacs,floatfix,preprint,a4paper]{revtex4-1}
\usepackage{amsmath,bm}
\usepackage{graphicx}
\usepackage{caption} 
\usepackage{subcaption}
\usepackage{commath}
\usepackage{epsfig}
\usepackage{cancel}
\usepackage{color,soul}
\usepackage{xcolor}
\usepackage{comment}
\usepackage[normalem]{ulem}
\usepackage[utf8]{inputenc}

\begin{document}

\title{Effect of quantizing magnetic field on the inner crusts of hot Neutron Stars}

\author{Rana Nandi$^1$, Somnath Mukhopadhyay$^2$ and Sarmistha Banik$^3$} 

\affiliation{$^1$ Polba Mahavidyalaya, Hooghly, West Bengal 712148, INDIA}
\affiliation{$^2$ NIT Tiruchirappalli, Tamil Nadu 620015, INDIA}
\affiliation{$^3$Dept. of Physics, BITS Pilani, Hyderabad Campus, Hyderabad-500078, INDIA. }
\email{nandi.rana@gmail.com, somnath@nitt.edu, sarmistha.banik@hyderabad.bits-pilani.ac.in}

\begin{abstract}
In the present work we study the effects of strongly quantizing magnetic fields and finite temperature on the properties of inner crusts of hot neutron stars. The inner crust of a neutron star 
contains neutron-rich nuclei arranged in a lattice and embedded in 
gases of free neutrons and electrons. 
We describe the system within the Wigner-Seitz (WS) cell approximation.
The nuclear energy is calculated using Skyrme model with SkM*
interaction. To isolate the properties of nuclei we follow the  subtraction procedure presented by Bonche, Levit and Vautherin, within
the Thomas-Fermi approximation. We obtain the equilibrium properties of inner crust for various  density, temperatures and magnetic fields by
minimizing the free energy of the WS cell satisfying the charge 
neutrality and $\beta-$equilibrium conditions.
We infer that at a fixed baryon density and temperature, strong quantizing magnetic field reduces the cell radii, neutron and proton numbers in the cell compared with the field free case. However, the nucleon number in the nucleus increases in presence of magnetic field.  The free energy per nucleon also decreases in magnetized inner crust. 
On the other hand, we find that finite temperature tends to smear out the effects of magnetic field. Our results can be important in
the context of $r-$process nucleosynthesis in the binary neutron
star mergers.

\end{abstract}

\maketitle

\section{Introduction}

Pulsars are highly magnetized rapidly rotating neutron stars having surface magnetic fields $\sim 10^{12}$ G. There is a class of neutron stars named magnetars consisting of Anomalous X-ray Pulsars (AXPs) and Soft Gamma-ray Repeaters (SGRs) with even higher surface magnetic field $\sim 10^{14}-10^{15}$ G \cite{Du1992,Vi2017}. Such intense magnetic field can be generated by dynamo mechanism in hot and newly born neutrons stars(NS) after core-collapse supernova \cite{Du1992,Th1995,Th1996}.  The maximum magnetic field that can exist inside the core is set by the Virial theorem \cite{Shapiro1983,Lai1991} and for a typical NS of mass $1.5M_\odot$ and radius 10 km the limiting field could be as high as $\sim 10^{18}$ G.

The properties of matter in such high magnetic field is an interesting field of research in theoretical astrophysics. The magnetic field is termed quantizing in the sense that the charged particles move in quantized orbits known as Landau levels in the direction perpendicular to the magnetic field \cite{ST68,LL77,CV77,Me92}. The charged particles become relativistic and Landau quantized when the cyclotron energy becomes comparable to the rest mass energy of the particle. Landau quantization changes the phase space and density of states and hence modifies the thermodynamic and transport properties of highly magnetized matter. Effects of Landau quantization have been studied extensively on the outer crust using magnetized Baym-Pethick-Sutherland (BPS) equation of state (EoS) \cite{Lai1991} and also using atomic mass models of Hartree-Fock-Bogoliubov \cite{Ch2012}. In the NS core Landau quantisation has been studied using relativistic mean field model \cite{Ferrer2010,Lopes2015, DebaPRL}. 

In this study we focus on the inner crust of a hot neutron star in the presence of a quantizing magnetic field and at finite temperature. The inner crust is composed of nuclei immersed in a neutron and electron gas. The matter is $\beta$-equilibrated and charge neutral and the nuclei are also in mechanical equilibrium with the neutron gas. In order to calculate the equilibrium properties of nuclei in the thermodynamic method, we follow the subtraction procedure of Bonche, Levit and Vautherin \cite{Bonche1984,Bonche1985,Suraud1987}. In this method, the nuclear properties are isolated from the nucleus plus gas phase using the subtraction procedure in a temperature-dependent Hartree-Fock theory \cite{Bonche1984,Bonche1985} as well as in zero and finite temperature extended TF calculations \cite{Suraud1987}. The subtraction procedure was extended to isolated nuclei embedded in a neutron gas \cite{De2001} and to nuclei in the inner crust at zero temperature \cite{Sil2002}. In a later work the subtraction procedure is also applied to nuclei in a strongly magnetized inner crust taking into effect of Landau quantization of degenerate electrons at zero temperature \cite{Nandi2011}. Here we extend the subtraction procedure to include strongly quantizing magnetic field and finite temperature effects using the finite-temperature magnetized TF formalism. This is relevant for inner crusts of newly
formed hot magnetars and binary neutron star mergers where the
crustal matter can be heated and the magnetic field can get amplified
during the merging process \cite{Ciolfi:2020cpf}.

\section{Formalism}

We study the  composition of the inner crust of NS for different temperatures and magnetic fields. 
The inner crust consists of nuclei arranged in a lattice and
immersed in free gases of electrons and neutrons. For finite
temperatures there might be free protons as well.
We employ the Winger-Seitz (WS) cell approximation where the lattice
is divided into spherical cells.
Each WS cell is considered to be charge neutral containing a nucleus at its center and the interaction between cells are ignored.
We assume that the matter is in $\beta$-equilibrium.
We further assume that the whole system is placed in a uniform magnetic field.
Strong magnetic field affects electrons as their motion in the plane
perpendicular to the field get quantized into Landau levels.
However, protons are affected only through the charge neutrality condition.

Due to the presence of nucleonic gases in the inner crust, the
spherical cell itself does not define a nucleus. How to isolate the properties of nuclei
in such a scenario was shown 
within both Hartree-Fock prescription \cite{Bonche1984, Bonche1985} and TF
formalism \cite{Suraud1987}. This is based on the fact that at a given
temperature and chemical potential there exists two solutions to the HF or TF equations, 
one solution corresponds to the nucleus in equilibrium with the nucleon gases while the other represents the nucleon gases alone.
The nucleus solution is obtained from the difference of two solutions.
In this work we adopt the TF formalism following Ref. \cite{Suraud1987}.

In order to obtain the thermodynamic properties of the system,
we need to minimize the total free energy of the WS cell 
under the conditions of charge neutrality and $\beta$-equilibrium. 
The relevant thermodynamic potential can be written as: 
\begin{equation}
\label{eq1}
\Omega = {\cal F}-\sum_q\mu_q A_q,
\end{equation}
where $q=(n,p)$ stands for neutrons and protons and $\mu_q$ are $A_q$ are the corresponding chemical potentials and numbers,
respectively.
The total free energy of the cell is a function of average baryon
number density ($\langle \rho\rangle$), temperature ($T$) and proton fraction ($Y_p$)
and can be expressed as: 
\begin{equation}
\label{eq:fe}
{\cal F}(\langle\rho \rangle,Y_p,T)=\int \left[{\cal H}(r)-Ts(r)
+ {\cal E}_c(r) +f_e(\rho_e)\right] d{\bf r} .
\end{equation}
Here ${\cal H}$ refers to the nuclear energy density excluding the Coulomb energy, $s$ is the entropy density of the nucleons, 
${\cal E}_c$ the Coulomb energy density of the system, $f_e$ is the free energy density of the electrons and $\rho_e$ is the
electron number density.

For the nuclear energy density ${\cal H}$, we adopt the Skyrme energy density functional which is written as 
\cite{Brack1985,RikovskaStone:2003bi}:
\begin{eqnarray}
\label{eq:ned}
{\cal H}(r)&=&\sum_q \frac{\hbar^2}{2m_q^*}\tau_q + 
\frac{1}{2}t_0\left[\left(1+\frac{x_0}{2}\right)\rho^2-\left(x_0+
\frac{1}{2}\right)\sum_q\rho^2_q\right]\nonumber\\
&&-\frac{1}{16}\left[t_2\left(1+\frac{x_2}{2}\right)
-3t_1\left(1+\frac{x_1}{2}\right)\right](\nabla \rho)^2\nonumber\\
&&-\frac{1}{16}\left[3t_1\left(x_1+\frac{1}{2}\right)
+t_2\left(x_2+\frac{1}{2}
\right)\right]\sum_q(\nabla\rho_q)^2\nonumber\\
&&+\frac{1}{12}t_3\rho^\alpha\left[\left(1+\frac{x_3}{2}\right)\rho^2
-\left(x_3+\frac{1}{2}\right)\sum_q \rho_q^2 \right].
\end{eqnarray}
We employ the SkM* interaction, for which the parameters $x_0$, $x_1$, $x_2$, $x_3$, $t_0$, $t_1$, $t_2$, $t_3$ and $\alpha$ can be found in \cite{Brack1985}.
In Eq. \ref{eq:ned}, $\rho=\rho_p+\rho_n$ and $m_q^*$ is the 
effective mass defined as: 
\begin{eqnarray}
\label{eq:effmass}
\frac{m}{m_q^*(r)}&=&1+\frac{m}{2\hbar^2}\left\{\left[t_1\left(1+
\frac{x_1}{2}\right)
+t_2\left(1+\frac{x_2}{2}\right)\right]\right.\rho \nonumber\\
&& +\left.\left[t_2\left(x_2+\frac{1}{2}\right)
-t_1\left(x_1+\frac{1}{2}
\right)\right] \rho_q\right\}.
\end{eqnarray}
The kinetic energy density $\tau_q$ takes the form in TF approximation as:
\begin{eqnarray}
\tau_q &=& \frac{3}{5} (3\pi^2)^{2/3}\rho_q^{5/3} \quad \text{for } T=0,\nonumber\\
 &=& \frac{1}{2\pi^2}\left( \frac{2 m_q^* T}{\hbar^2} \right)^{5/2} J_{3/2}(\eta_q) \quad \text{for } T\neq 0,
\label{eq:ked}
\end{eqnarray}
where $\eta_q$ is the fugacity that is obtained from the
chemical potential and single particle potential of nucleons
$V_q$ as:
\begin{equation}
\label{eq:fugacity}   
\eta_q(r)=(\mu_q-V_q(r))/T
\end{equation}
The number density of nucleons is given by  
\begin{equation}
\label{eq:nd}
\rho_q(r)= \frac{1}{2\pi^2}\left( \frac{2 m_q^* T}{\hbar^2} \right)^{3/2} J_{1/2}(\eta_q).
\end{equation}
The functions $J_k(\eta_q)$ appearing in Eqs. (\ref{eq:ked})
 and (\ref{eq:nd}) are Fermi integrals:
\begin{equation}
\label{eq:fi}
J_k(\eta _q) = \int_0^\infty \frac{x^k}{\exp(x-\eta_q)+1}\,dx
\quad .
\end{equation}
The entropy density of the nucleons is related to fugacity and number density as
\begin{equation}
\label{eq:ent}
s(r)= \sum_q [(5/3)J_{3/2}(\eta_q)/J_{1/2}(\eta_q) - \eta_q] 
\, \rho_q .
\end{equation}
The third term in the integrand of Eq. 2 is the Coulomb energy density of the charged particles and is given by
\begin{equation}
\label{eq10}
{\cal E}_c(r)=\frac{1}{2} (\rho_p(r)-\rho_e) \int
(\rho_p(r^{\prime})-\rho_e)\frac{e^2}{\mid{\bf r}-
{\bf r^{\prime}}\mid} d{\bf r^{\prime}}.
\end{equation}
We neglect the exchange term for the Coulomb energy density.

The density profiles of nucleus+gas (NG) phase $\rho_{\rm NG}^q$
and gas (G) phase $\rho_{\rm G}^q$  are obtained from the
variational equations:
\begin{equation}
\label{eq:var_ng}
\frac{\delta \Omega_{\rm NG}}{\delta \rho_{\rm NG}^q}=0,
\end{equation}
\begin{equation}
\frac{\delta \Omega_{\rm G}}{\delta \rho_{\rm G}^q}=0,
\end{equation}
where $\Omega_{\rm NG}$ and $\Omega_{\rm G}$ are the thermodynamic
potentials of the corresponding phases.
This results in the following coupled equations \cite{Sil2002, Nandi2011}
\begin{eqnarray}
T\eta_{\rm NG}^q(rn)+V_{\rm NG}^q + V_{\rm NG}^c(\rho_{\rm NG}^p,\rho_e) &=&\mu_q\, ,\\
T\eta_{\rm G}^q(r)+V_{\rm G}^q + V_{\rm G}^c(\rho_{\rm NG}^p,\rho_e)&=&\mu_q\, .
\label{eq:cp_t}
\end{eqnarray}
At $T=0$ these two equations simplify to:
\begin{eqnarray}
(3\pi^2)^{2/3}\frac{\hbar^2}{2m_q^*}(\rho_{\rm NG}^q)^{2/3} + V_{\rm NG}^q + V_{\rm NG}^c(\rho_{\rm NG}^p,\rho_e) &=& \mu_q\, ,\\
(3\pi^2)^{2/3}\frac{\hbar^2}{2m_q^*}(\rho_{\rm NG}^q)^{2/3} + V_{\rm G}^q + V_{\rm G}^c(\rho_{\rm NG}^p,\rho_e) &=& \mu_q\, ,
\end{eqnarray}
where $V_{\rm NG}^q$ and $V_{\rm G}^q$ are the nuclear part of the
single particle potentials  in the nucleus+gas and gas phases,
respectively and $V_{\rm NG}^c$ and $V_{\rm G}^c$ correspond to the Coulomb part of the single particle potential for two phases and are given by the same expression as: \cite{Sil2002, Nandi2011}
\begin{equation}
\label{eq:sp_c}
 V^c(r) = \int\left[\rho_{\rm NG}^p(r') - \rho_e\right] \frac{e^2}{\mid{{\bf r}-{\bf r'}}\mid} d{\bf r'}~.
\end{equation}
The average chemical potential for the $q-$th nucleon is given by
\begin{equation}
\label{eq:muq_av}
\mu_q=\frac{1}{A_q}\int \left[ T\eta_{\rm NG}^q(r)+V_{\rm NG}^q(r)
+V_{\rm NG}^c(r)\right] \rho_{\rm NG}^q(r) d{\bf r} ,
\end{equation}
where $A_q$ refers to $N_{\rm cell}$ or  $Z_{\rm cell}$,
where $N_{\rm cell}$ and $Z_{\rm cell}$ are neutron and proton numbers
in the WS cell, respectively, which can easily be obtained from the
average baryon density $\langle \rho \rangle$, proton fraction $Y_p$
and the cell radius $R_c$ as:
\begin{equation}
    Z_{\rm cell} = Y_p\,\langle \rho \rangle\, V_{\rm cell}, \qquad 
    N_{\rm cell} = (1-Y_p)\,\langle \rho \rangle \,V_{\rm cell}\, ,
   \quad \text{ and } \quad  V_{\rm cell}=\frac{4}{3}\pi R_c^3\, ,
   \label{eq:nz_cell}
\end{equation}
where $V_{\rm cell}$ is the volume of the cell. The 
total number of nucleons in the cell is
$A_{\rm cell}=N_{\rm cell}+Z_{\rm cell}$.

Finally, number of neutrons ($N$) and protons ($Z$) in a nucleus are obtained from the subtracted densities as
\begin{eqnarray}
\label{eq:nz}
Z&=&\int \left[\rho_p^{\rm NG}(r)-\rho_p^{\rm G}(r)\right]d{\bf r}~,\nonumber\\
N&=&\int \left[\rho_n^{\rm NG}(r)-\rho_n^{\rm G}(r)\right]d{\bf r}~,
\end{eqnarray}
so that the mass number of the nucleus is $A=N+Z$.

We assume the matter to be in $\beta$-equilibrium, the chemical potential $\mu$ of the species are constrained by the relation:
\begin{equation}
\label{eq:beqm}
\mu_e = \mu_n - \mu_p + \Delta m~.
\end{equation}
where $\Delta m$ is the mass difference between neutrons and protons.

\subsection{Effect of Quantizing Magnetic field}

The properties of inner crust are significantly influenced in the presence of strong quantizing magnetic field. We assume a uniform magnetic field $(0,0,B)$ in the crust. The motion of electrons is Landau quantized in the plane perpendicular to the magnetic field, which indirectly affects the properties of protons in the charge neutral WS cells and hence the nuclei. For $B > B_c =m_e^2/e \simeq 4.414 \times 10^{13}$ G (with $\hbar=c=1$), the transverse motion of the electrons becomes relativistic \cite{Lai1991}. The quantized energy levels of the electrons with momentum $p_z$ for $\nu$-th Landau level is given by (with $B_*=B/B_c$)
\begin{equation}
\label{eq:ee}
E_e(\nu, p_z)= \left[p_z^2 + m_e^2 (1 + 2 \nu B_{*})\right]^{1/2}
\end{equation}
The number density of electrons in the magnetic field at finite temperature can be written as \cite{Lai2001}:
\begin{equation}
\label{eq:nde_T}
\rho_e = \frac{m_e^2B_*}{4\pi^2}\sum_{\nu=0}^\infty g_\nu \int_{-\infty}^{\infty}f\, dp_z
\end{equation}
where
\begin{equation}
    f=\frac{1}{1 + e^{\beta(E_e-\mu_e)}} \label{eq:fd}
\end{equation}
is the Fermi-Dirac distribution function with $\beta=1/(k_BT)$, $\mu_e$ is the electron chemical potential, $g_\nu$ is the spin degeneracy of the Landau level ( $g_0=1$ and $g_\nu=2$ for $\nu \geq 1$). 
At $T=0$ the above expression simplifies to:
\begin{equation}
    \rho_e = \frac{m_e^2B_*}{2\pi^2}\sum_{\nu=0}^{\nu_{\rm max}} g_\nu\, p_f(\nu)\, ,
    \label{eq:nde_T0}
\end{equation}
where $p_f(\nu)$ is the maximum $z-$component of electron momentum 
and is related to the chemical potential as:
\begin{equation}
    p_f^2(\nu) + m_e^2(1+2\nu B_*) = \mu_e^2\, ,
    \label{eq:mue}
\end{equation}
and $\nu_{\rm max}$ is highest Landau level that can be populated
for a given $B_*$ and $\mu_e$ and is obtained by noting that $p_f^2(\nu)\geq0$:
\begin{eqnarray}
 \nu_{\rm max}=\frac{\mu_e ^2-m_e^2}{2m_e^2B_*}~.
 \label{eq:numax_T0}
\end{eqnarray}
The energy density of electrons is obtained from
\begin{eqnarray}
\label{eq:ede_T}
\varepsilon_e &=&\frac{m_e^2B_*}{4\pi^2}\sum_{\nu=0}^{\infty}g_\nu \int_{-\infty}^{\infty}f E_e(\nu,p_z)\,dp_z\qquad\text{for}\quad T\neq 0\, ,\\
             &=& \frac{m_e^2B_*}{2\pi^2}\sum_{\nu=0}^{\infty}g_\nu
\int_{0}^{p_f(\nu)} E_e(\nu,p_z)\,dp_z\qquad\text{for}\quad T=0\,.
\label{eq:ede_T0}
\end{eqnarray}

\section{Results}

We present the properties of matter in the inner crusts
of neutron stars under the influence of strong magnetic fields and finite temperature. In particular, we
study the effects of Landau quantization of electrons on the number of nucleons in the NS crusts. For this purpose we 
consider magnetic fields $B_*$=10, 100, 10$^3$ and $10^4$.
The effect of magnetic field is noticeable if only first few levels
are populated.
The highest Landau level $\nu_m$ populated at a given $B_*$ can be
calculated from Eq. \ref{eq:numax_T0} for $T=0$. On the other 
hand, one needs to evaluate infinite sums over Landau levels
to calculate number and energy densities of electrons (Eq.
\ref{eq:nde_T} and \ref{eq:ede_T}), for $T\neq0$.
In practice, only finite number of levels contribute when the magnetic
field is strong. In the numerical implementation we set 
the Fermi function  (Eq. \ref{eq:fd}) to zero when 
$\beta(E_e(\nu,p_z)-\mu_e)\geq30$.
This condition along with Eq. \ref{eq:ee} leads to the highest
Landau level as:
\begin{equation}
    \nu_{\rm max} = \frac{1}{2B_*}\left[\left(\frac{30\,T+\mu_e}{m_e} \right)^2 -1 \right]\, .
\end{equation}
Eq. \ref{eq:numax_T0} is recovered by putting $T=0$.

We demonstrate the results only for $B_*=10^4$ when
only the zeroth Landau level is populated (at least for $T=0$).
For lower values of $B_*$ there is no
visible effect as several Landau levels are populated by electrons.  For comparison, we also show the results for $B_*=0$ cases. We consider the crust temperature in the range of $0-4$ MeV. 

\begin{figure}
\centering
\begin{tabular}{cc}
\includegraphics[width=0.45\textwidth]{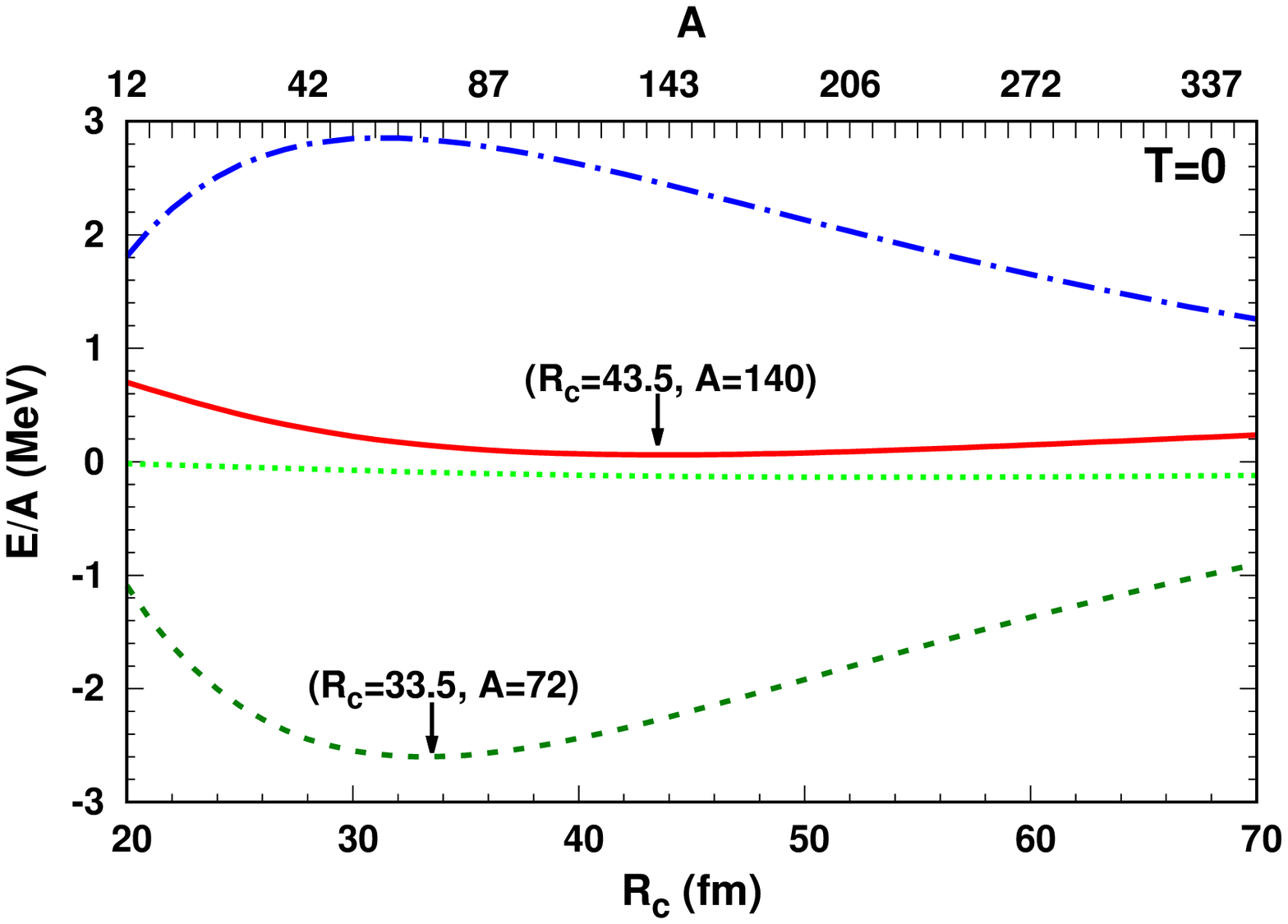} &
\includegraphics[width=0.45\textwidth]{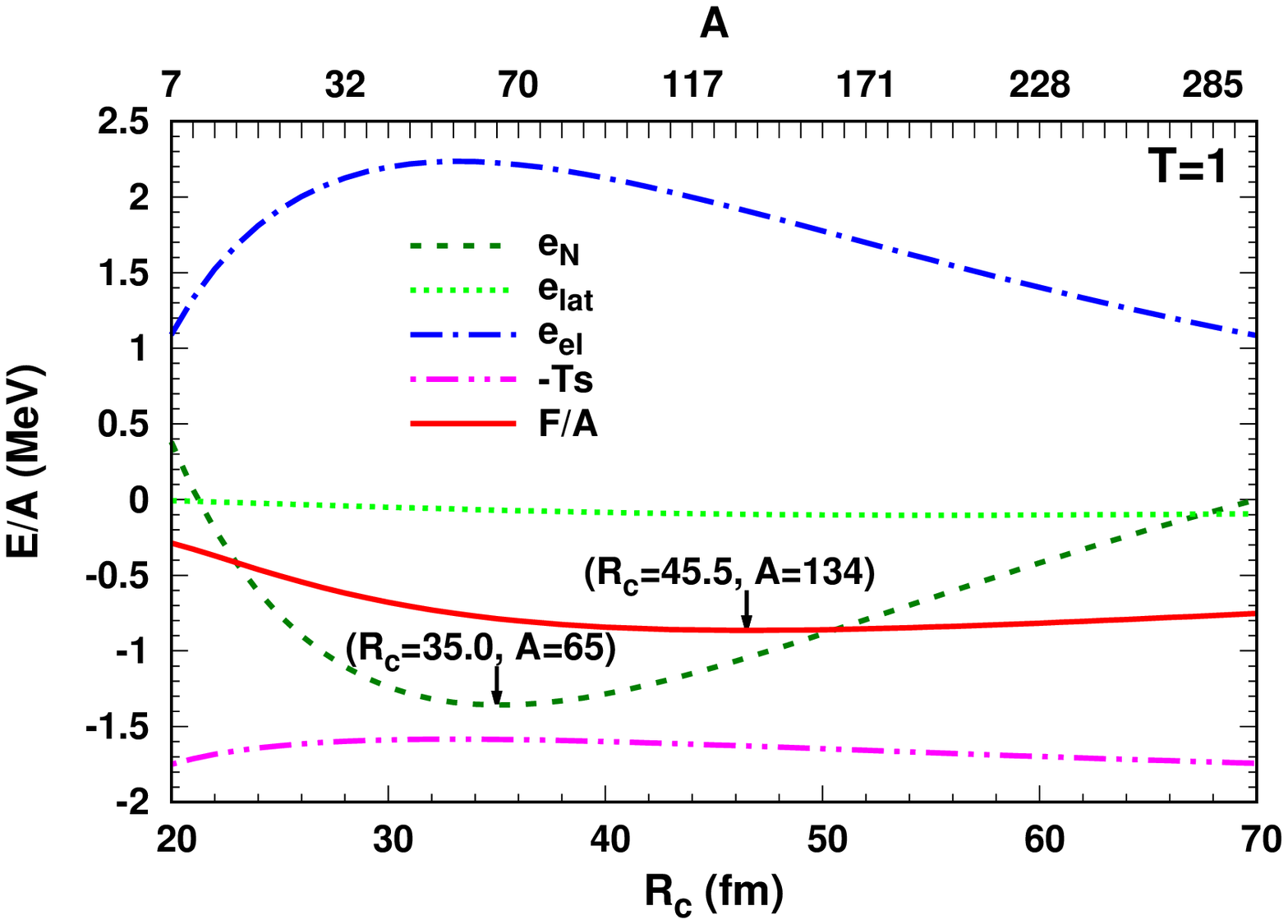} \\
\includegraphics[width=0.45\textwidth]{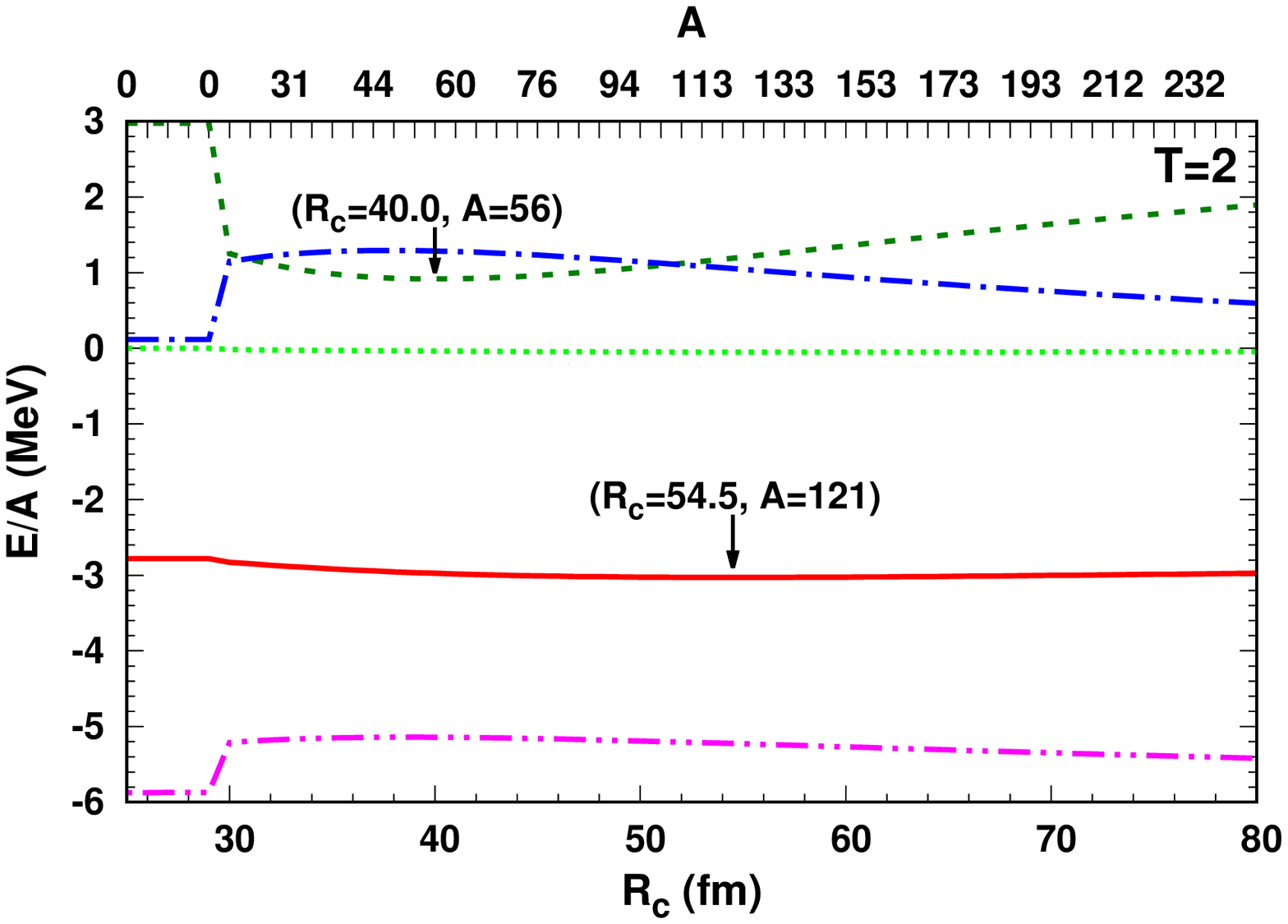} &
\includegraphics[width=0.45\textwidth]{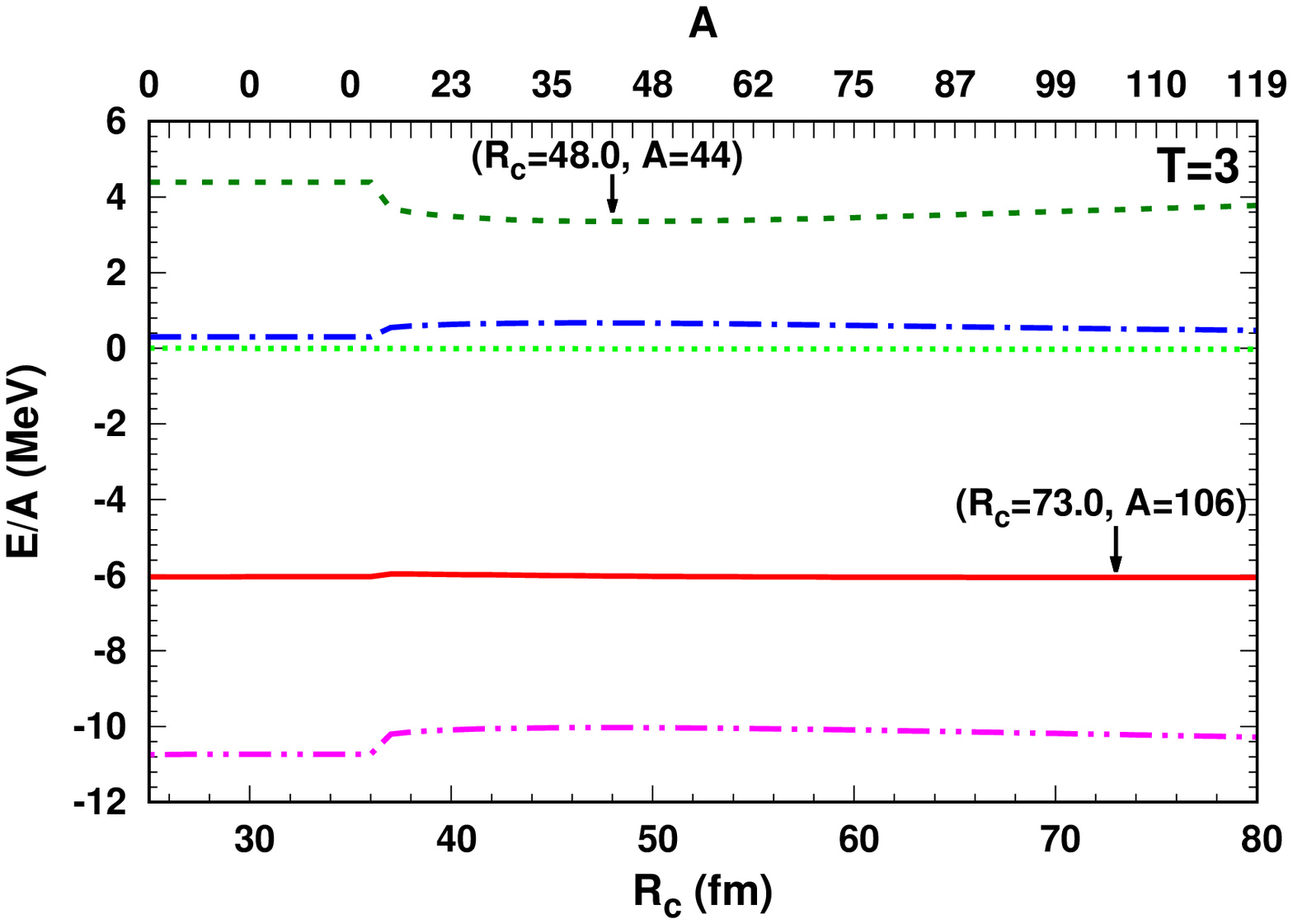}
\end{tabular}
\caption{Different contributions to the free energy per nucleon as
a function of $R_c$ and $A$ (upper x-axis) for $T=0-3$ MeV, $B=0$ and 
$\langle \rho \rangle =0.001$ fm$^{-3}$.}
\label{fig:energies_B0}
\end{figure}
In order to obtain the equilibrium configuration for a given average number density ($\langle \rho\rangle $), temperature ($T$) and
magnetic field ($B_*$), we minimize the free energy of the WS cell
by varying $Y_p$ and $R_c$ while satisfying the conditions of charge
neutrality and $\beta-$equilibrium.
The free energy minimum is governed by the contribution of different
terms:
\begin{equation}
    F/A_{\rm cell} = e_N + e_{\rm lat} + e_{\rm el} - Ts\, ,
    \label{eq:facell}
\end{equation}
where $e_N$ is nuclear energy per nucleon including the Coulomb
energy of protons, $e_{\rm lat}$ is the lattice energy per nucleon
which consists of electron-proton and electron-electron Coulomb
energies, $e_{\rm el}$ is the kinetic energy of electrons per nucleon
and $s$ is the entropy per nucleon including the entropy of electrons.
In Fig. \ref{fig:energies_B0}, we show the variation of all the  
components of free energy per nucleon with the cell radius or
equivalently $A_{\rm cell}$ 
(since, $A_{\rm cell}=V_{\rm cell}\langle\rho\rangle$) 
at a density $\langle\rho\rangle=0.001$ fm$^{-3}$ and at
different temperatures ($T=0-3$ MeV) for the non-magnetic case.
The upper x-axis presents the corresponding mass number of nuclei 
obtained from the subtracted densities (see Eq. \ref{eq:nz}).
It is seen from the figure that for each $T$, $e_N$ has a minimum
at a certain $R_c$. Since, the nuclear mass number $A$
(see upper x-axis) increases monotonically with $R_c$, the minimum in
$e_N$ corresponds to the nucleus for which the nuclear binding energy
per nucleon is minimum. These nuclei are very neutron-rich and
get smaller with increasing $T$. The free energy minima are at 
larger $R_c$ as they originate from the competition between 
different terms, but mostly dominated by $e_N$ and $e_{\rm el}$
and also $Ts$ at finite temperature. An interesting feature appears
at $T=2$ MeV, where a sudden change in each component of free energy
is observed at $R_c= 30$ fm. The upper x-axis reveals that for 
$R_c< 30$ fm $A=0$, which essentially means that no nucleus exists
till that $R_c$ and the matter is completely in the gas state,
instead. However, the complete gas state has higher free energy than
the nucleus+gas state  for which the free energy is minimum at
$R_c=54.5$ fm. 
Similar behaviour is found at $T=3$ MeV, where nuclei come into the 
picture only after $R_c=36$ fm. Still we find that the free energy
minimum corresponds to nuclear+gas solution at $R_c=73$ fm. The 
free energy difference between the complete gas state and the 
nuclear+gas state decreases with $T$. At $T=2$ MeV the minimum values
of free energies in the two states are $-2.780$ and $-3.029$ MeV.
Whereas, for $T=3$ MeV the corresponding values are $-6.038$ and
$-6.057$ MeV. For $T=4$ MeV or higher (not shown here) only the gas
solution exists. We can say that the critical temperature of liquid-gas
phase transition lies between $T=3$ and 4 MeV at this density.
This agrees with the results of other studies \cite{Nandi:2016ylp}.

\begin{figure}
\centering
\begin{tabular}{cc}
\includegraphics[width=0.45\textwidth]{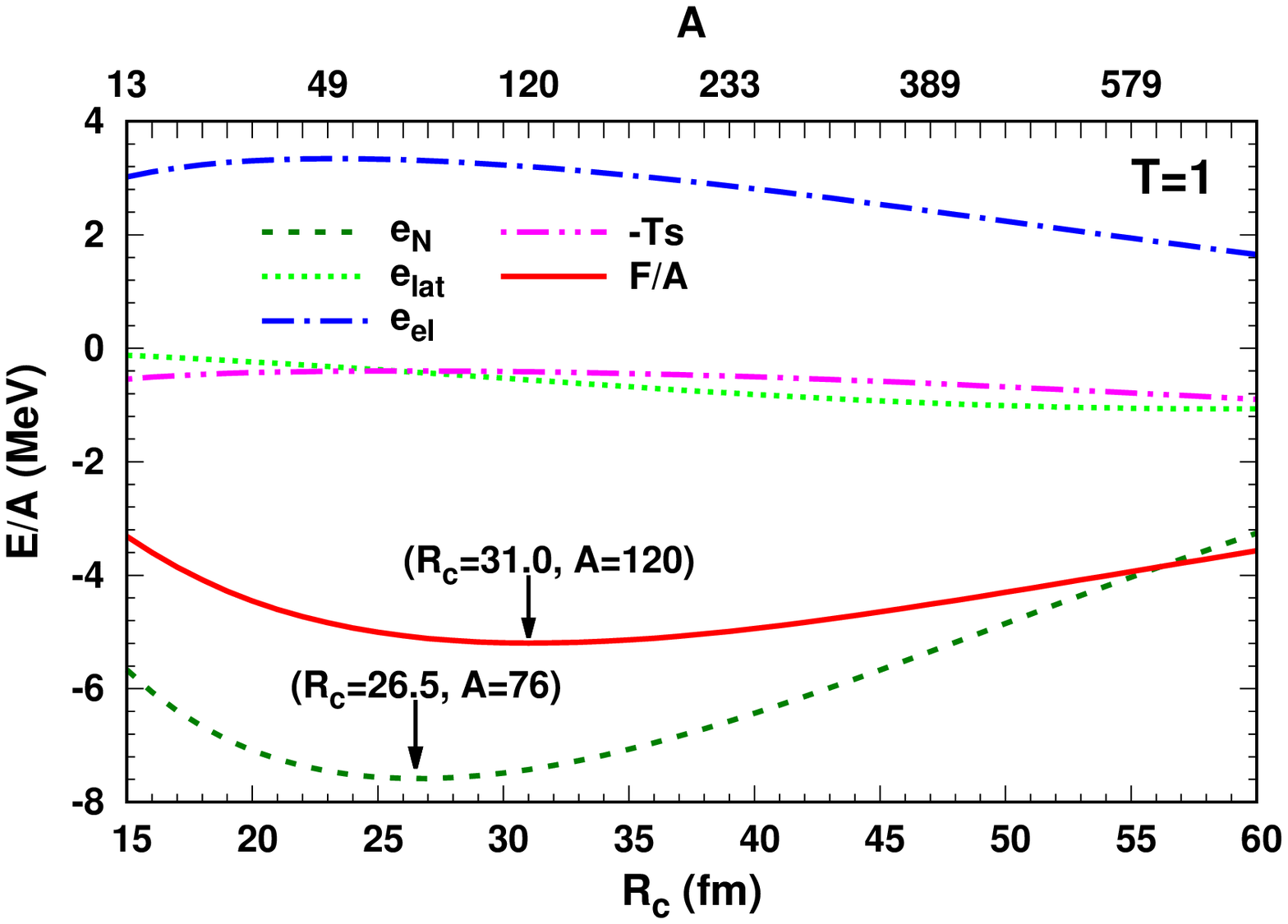} &
\includegraphics[width=0.45\textwidth]{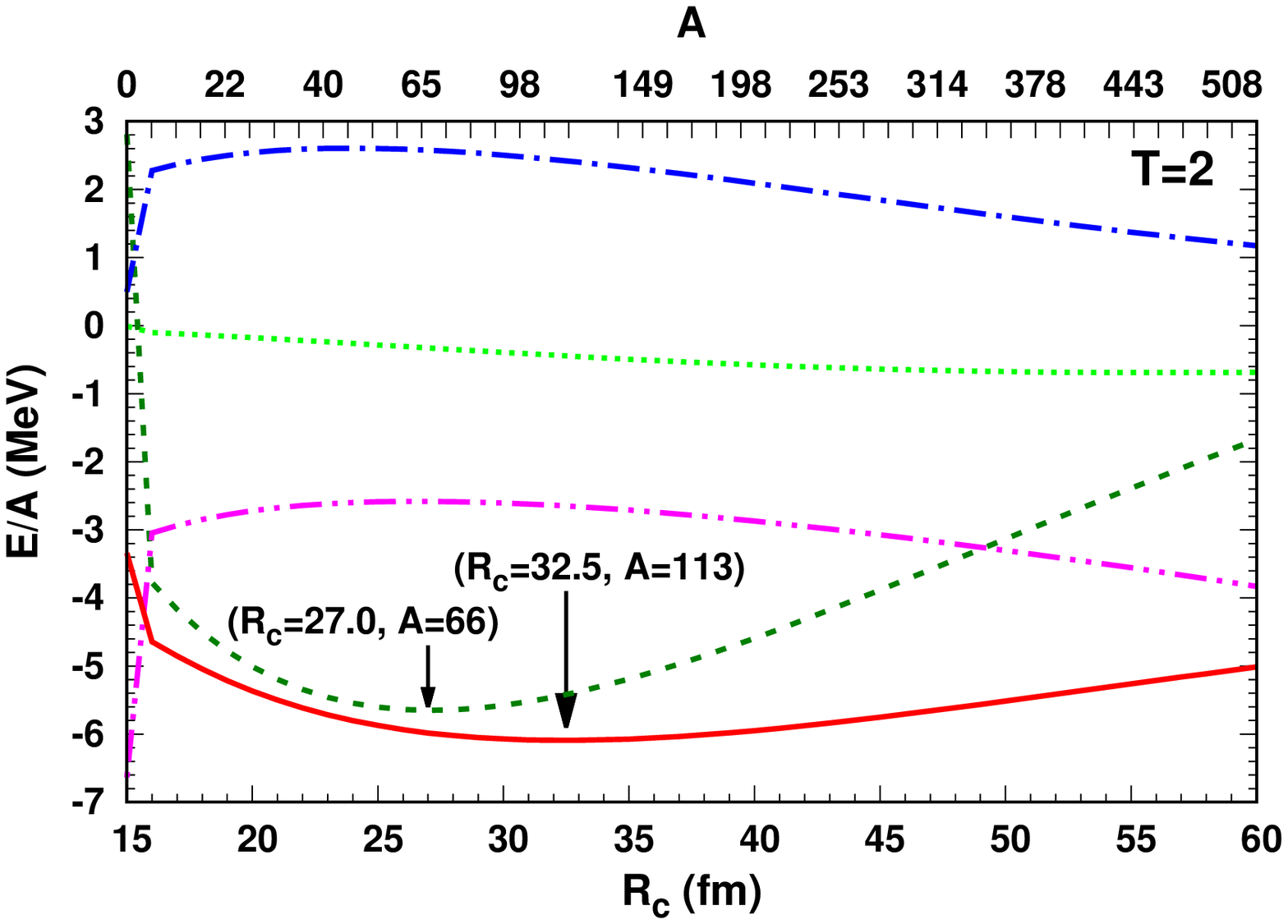} \\
\includegraphics[width=0.45\textwidth]{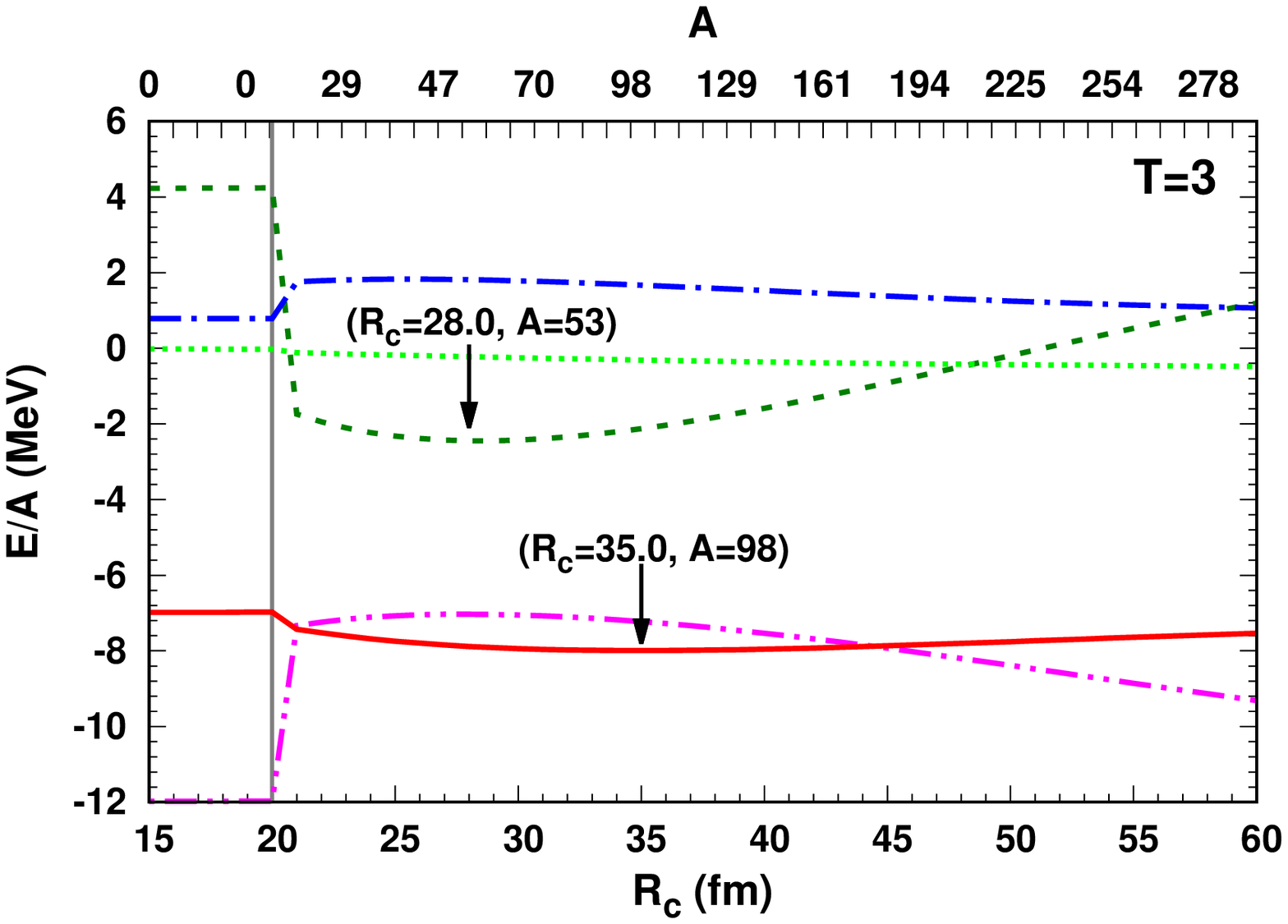} &
\includegraphics[width=0.45\textwidth]{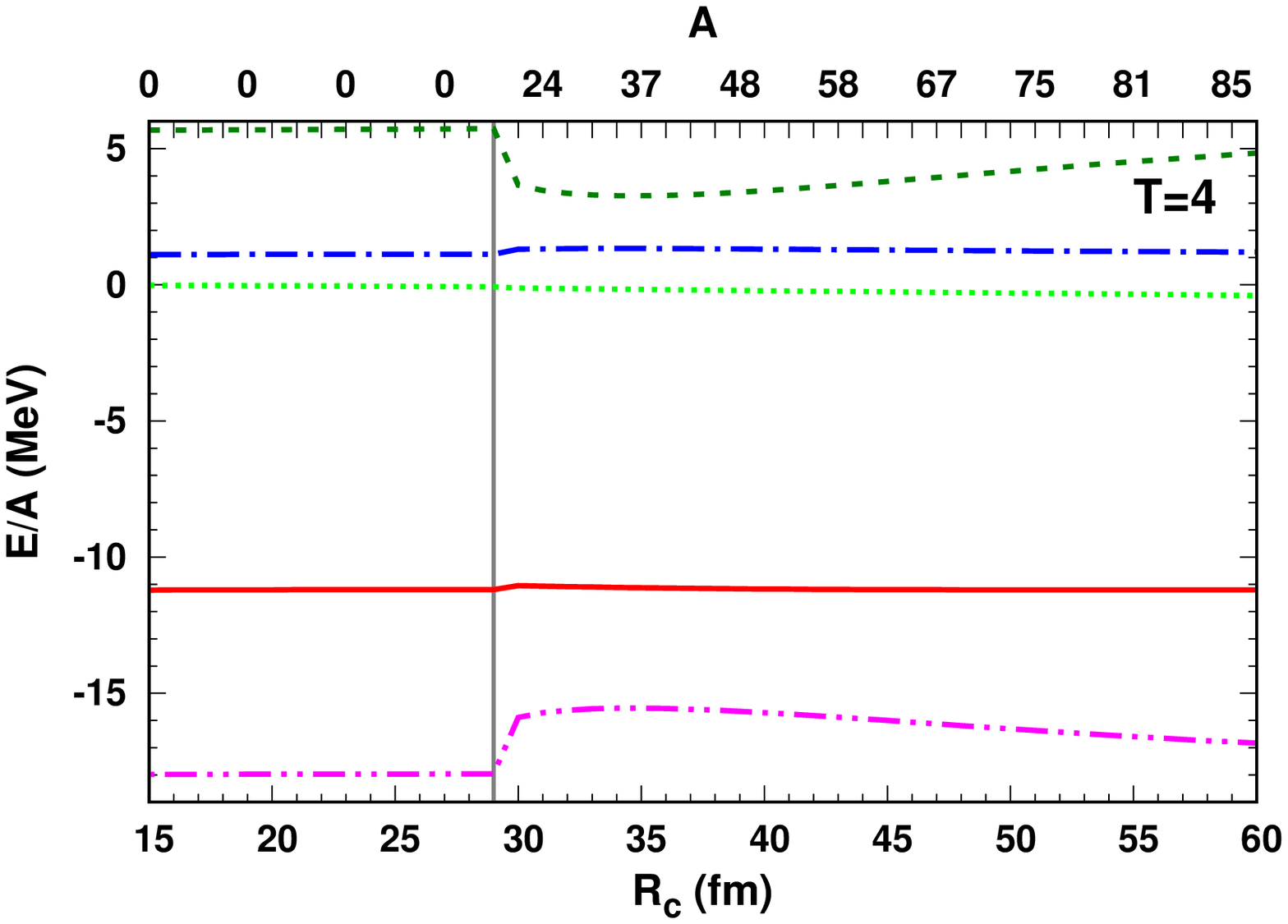}
\end{tabular}
\caption{Same as Fig. \ref{fig:energies_B0} but for $T=1-3$ MeV, and $B*=10^4$.}
\label{fig:energies_B10k}
\end{figure}
In Fig. \ref{fig:energies_B10k} we present the variation of 
free energy and its different components with $R_c$ as in
Fig. \ref{fig:energies_B0}, but now in the presence of magnetic field with $B_*=10^4$. We do not show the plot for $T=0$, since there are no drip neutrons  at $\langle \rho \rangle=0.001$ fm$^{-3}$,
unlike the non-magnetic case. This is consistent
with the earlier studies where it was found that strong magnetic
fields shift the neutron drip-point to higher densities
\cite{Nandi:2010fp}. Similar to the non-magnetic case we find that
with increasing $T$ the cell radius corresponding to both the $e_N$
minimum and the free energy minimum increase, whereas the mass
number of the equilibrium nucleus decreases. At $T=3$, matter is
uniform till $R_c=20$ fm and beyond that nucleus appears with free
energy minimum at $R_c=35$ fm, where $A=98$. On the other hand,
at $T=4$ uniform gas solution extends up to 29 fm and has lower
free energy than the solution with nucleus.

Comparison between Fig. \ref{fig:energies_B0} and
\ref{fig:energies_B10k} shows that when $B_*=10^4$ the minima in $e_N$
and free energy are obtained at lower $R_c$. It can also be
seen that $e_N$ plays more dominant role in deciding the free 
energy minimum. As a result, the $R_c$ corresponding to the free energy
minima are not very far from that of $e_N$, unlike non-magnetic
scenario. Comparing these two 
figures we also observe that a strong magnetic field affects all the
components of free energy. But the effect is found to be  most
significant on $e_N$ which gets appreciably reduced when magnetic field
of $B_*=10^4$ is applied.
To understand the reason we note that for a given
$\langle \rho \rangle$ and $Y_p$ although the electron density is the
same for both the scenarios (since $\rho_e=Y_p\langle \rho \rangle$),
the electron chemical potential $\mu_e$ decreases significantly.
For instance, in the present example of $\langle \rho \rangle=0.001$
fm$^{-3}$, the $e_N$ minimum at $T=1$ MeV for the non-magnetic
case corresponds to $R_c=35$ fm, $Y_p=0.102$,
$\rho_e=1.02\times10^{-4}$ fm$^{-3}$ and $\mu_e=28$ MeV.
When the magnetic field of magnitude $B_*=10^4$ is switched on
keeping the values of $\langle\rho\rangle$, $T$, $R_c$ and $Y_p$ same
we find that $\mu_e$ reduces from 28 to 6 MeV, whereas $\mu_n$ and
$\mu_p$ remain unaltered. The decrease in $\mu_e$ is caused by the
phase space modification of electrons in strongly quantizing magnetic
field. For $B_*=10^4$ only the first Landau level ($\nu=0$) gets
populated leading to a smaller value of $\mu_e$ than the non-magnetic
case, for a given $\rho_e$ (see eqs. \ref{eq:nde_T0} and \ref{eq:mue}).
Therefore, to achieve the $\beta-$equilibrium
one needs to increase $Y_p$, which in turn increases $\rho_e$. 
For the present example, $Y_p$ has to be increased from $0.102$ to $0.321$ with corresponding increase in $\rho_e$ from $1.02\times10^{-4}$
to $3.21\times10^{-4}$ fm$^{-3}$ to satisfy the 
$\beta-$equilibrium condition. 
Although more electrons in the system increases their
kinetic energy to some extent but the nuclear energy and therefore
the free energy is greatly reduced due to the reduction in the Coulomb
energy. This can also be observed by noting the values of $e_N$ and
$e_{\rm el}$ from Figs. \ref{fig:energies_B0} and \ref{fig:energies_B10k}. In other words, strong magnetic field
increases the binding energy of the system. This is also the reason
why the solution with nuclei survives up to a higher temperature in
the presence of strong magnetic fields.

\begin{figure}[ht]
\begin{tabular}{cc}
\includegraphics[width=0.55\textwidth, angle=-90]{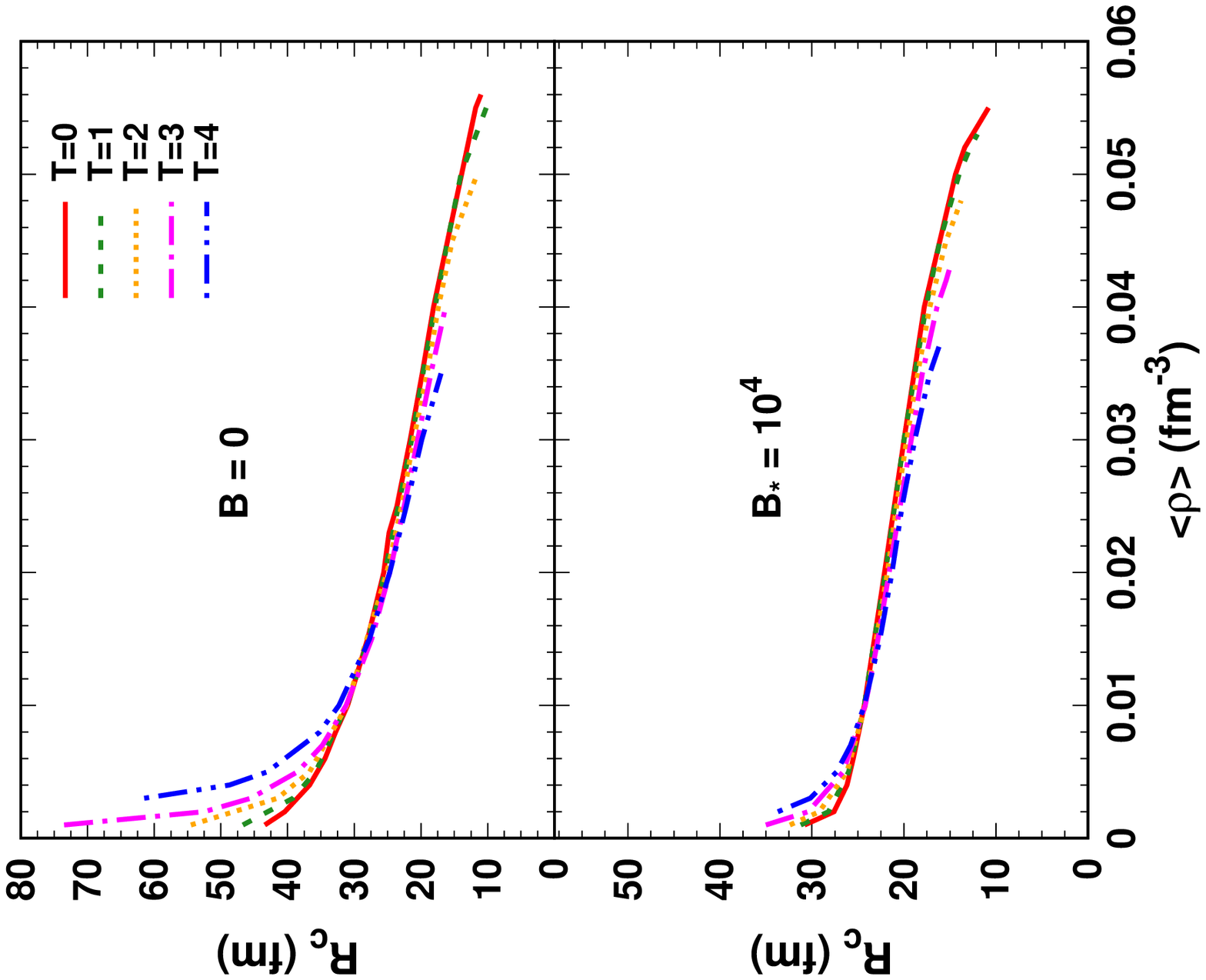} &
\includegraphics[width=0.55\textwidth, angle=-90]{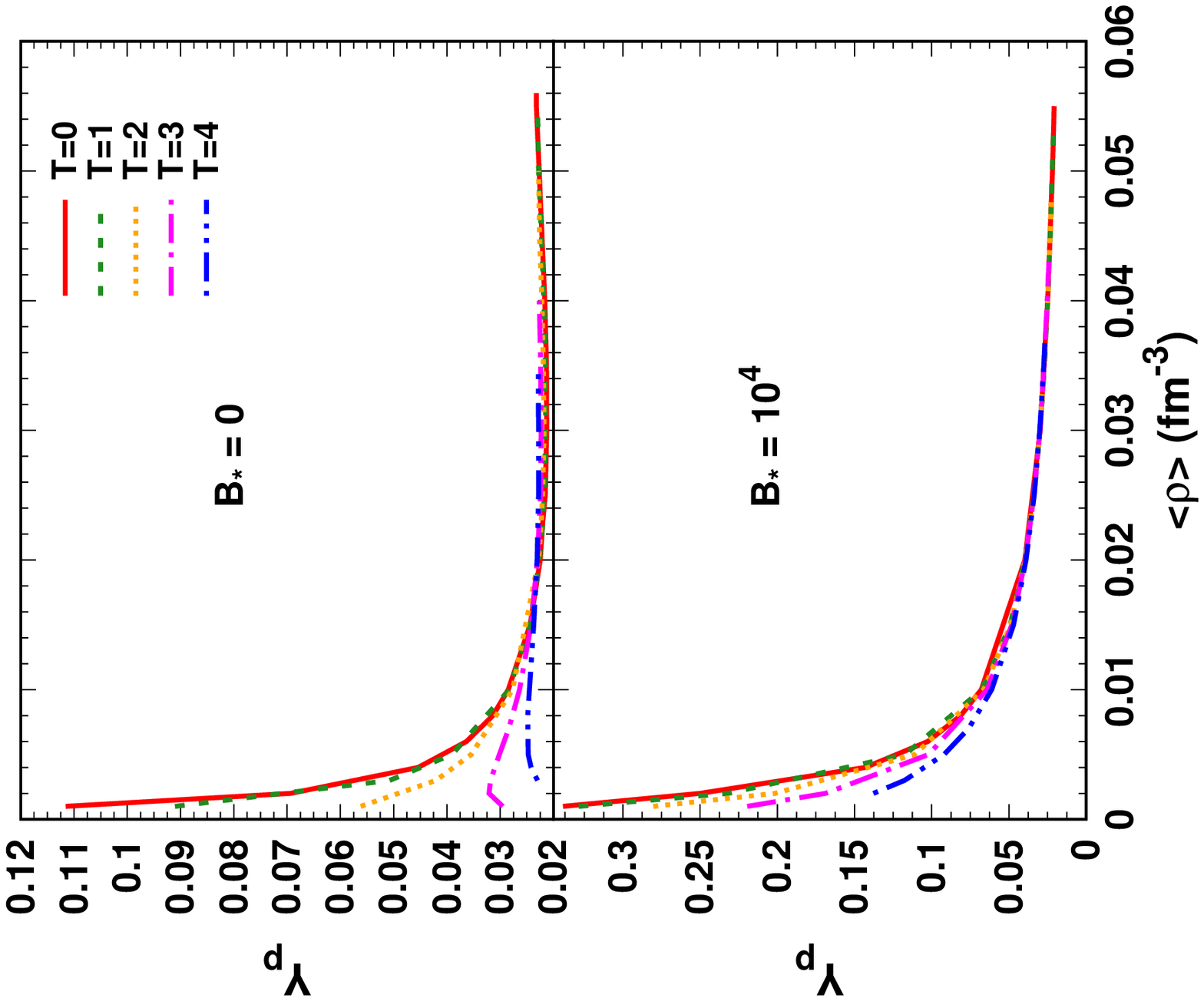}  \\
\end{tabular}
\caption{WS cell radius $R_c$ (left panel) and proton fraction $Y_p$
(right panel) as a function of average baryon density for $T=0-4$ MeV 
with and without quantizing magnetic field.}
\label{fig:Rc&Yp}
\end{figure}
The upper left panel of Fig. \ref{fig:Rc&Yp} shows the WS cell radii
corresponding to free energy minima as a function of number density
of nucleons for a range of temperature ($0-4$ MeV). The size of the
cell always decreases with increasing  baryon density.
In fact, the WS cell size is quite large at very low density, as was
also shown in Fig. \ref{fig:energies_B0}, it shoots up at
$\langle\rho\rangle=0.001$ fm$^{-3}$ for $T=3$ MeV compared to zero
temperature case. For $T=4$ MeV, however, the matter becomes completely
uniform at $\langle\rho\rangle\lesssim0.002$ fm$^{-3}$.
On the high density side ($\langle\rho\rangle\gtrsim 0.015$
fm$^{-3}$), at a given density the WS cell size is smaller for high
temperature. Also, as expected, the nuclei dissolve into uniform
matter at a relatively lower density for hot inner crust. 
In the  presence of strongly quantizing magnetic field, the nature of the curves in the lower left panel of  Fig \ref{fig:Rc&Yp}
remains more or less similar.  We see that for low densities
($\langle\rho\rangle\lesssim0.01$ fm$^{-3}$) finite temperature causes the WS cell radii to increase, but the cell size is not as large as $B=0$ case. At a fixed temperature magnetic field reduces
the cell
size. This is again because at very high magnetic field the
Coulomb interaction becomes much more efficient in increasing the
nuclear binding energy which then plays the most dominant
role in deciding the free energy minimum and thereby leading to the
reduction in the cell radius than that of the field-free case.
With rise in temperature more and more neutrons become unbound
and drip out of the nuclei. As a result, the nuclear
binding energy decreases and cell size increases.

In the right panel of Fig. \ref{fig:Rc&Yp} we display the dependence
of proton fraction on average baryon density for $T=0-4$ MeV and
with (lower panel) and without (upper panel) quantizing magnetic field.
We note that proton fraction decreases with temperature 
for both the non-magnetic and magnetic cases. 
It is also observed that magnetic field of strength $B_*=10^4$ enhances
the proton fraction significantly, especially at low densities. This
is because strong magnetic field reduces electron chemical potential and
therefore higher proton fraction is needed to maintain the
$\beta-$equilibrium. We have already discussed it in detail in
connection with Fig. \ref{fig:energies_B0} and Fig. \ref{fig:energies_B10k}.

\begin{figure}
    \centering
    \begin{tabular}{cc}
    \includegraphics[width=0.55\textwidth,angle=-90]{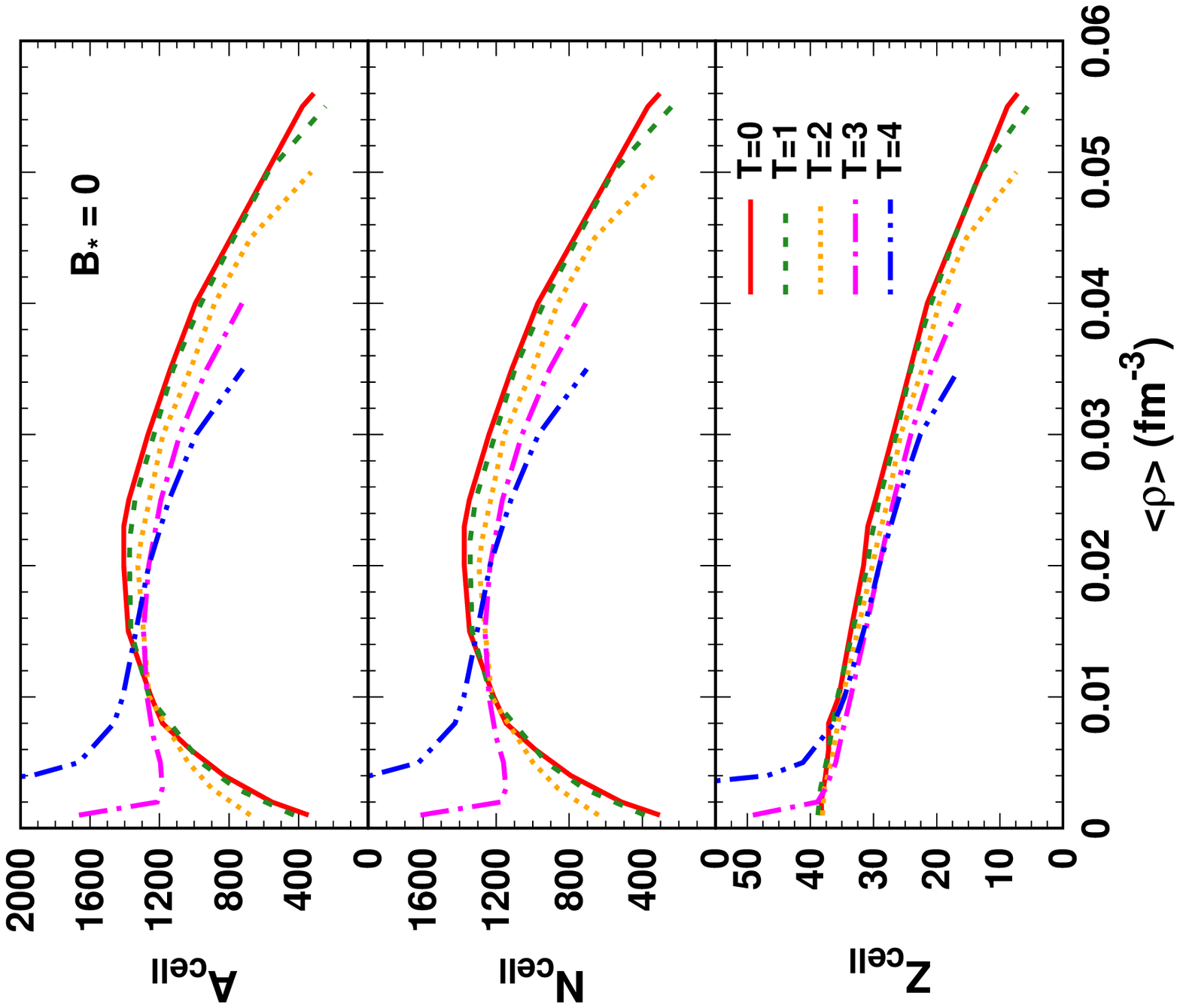} &
    \includegraphics[width=0.55\textwidth,angle=-90]{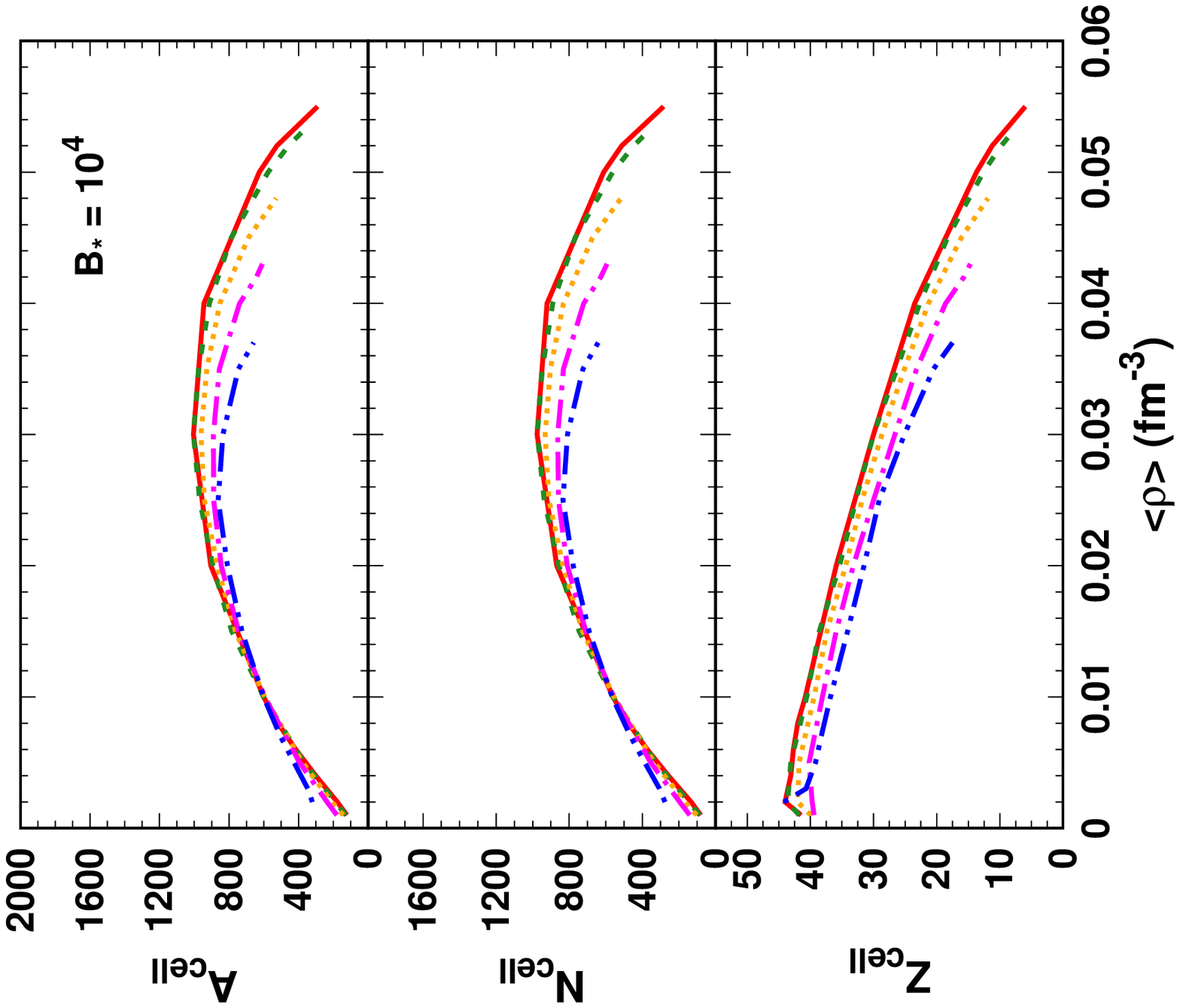}
    \end{tabular}
    \caption{Total number of neutrons ($N_{\rm cell}$), protons
    ($Z_{\rm cell}$) and $A_{\rm cell}=N_{\rm cell} + Z_{\rm cell}$ in the cell as a function of average baryon density and
    temperature for $B_*=0$ (left panel) and $B_*=10^4$
    (right panel).}
    \label{fig:Ncell}
\end{figure}
The top panels of Figs. \ref{fig:Ncell} show the total number of
nucleons $A_{\rm cell}(=4/3\pi R_c^3\langle\rho\rangle)$ inside the WS
cell, as a function of baryon density in non-magnetic (left panel) and
magnetic (right panel) NS inner crust, respectively. In both the cases,
the number grows to a maximum before falling down at higher densities.
These findings are consistent with earlier studies performed at $T=0$ \cite{Nandi2011,NV,Cheng1997}. 
In $B_*=0$ case, at very low density, the number shoots up for  higher
temperature. This is the consequence of high values of $R_c$ at low
densities, as noted earlier. At higher densities, $A_{\rm cell}$ is
maximum at $T=0$ and decreases with temperature for both the non-magnetic and magnetic cases.
However, for $B_* = 10^4$, the cell can accommodate a lesser number of
nucleons as evident from Fig. \ref{fig:Ncell}.

In the middle and bottom panels of Fig. \ref{fig:Ncell} we plot the individual number of  neutrons and protons in the WS cell, in the absence and presence of magnetic field ($B_*=10^4$). The proton number decreases monotonically with higher baryon density, whereas the total number of neutrons rises with number density, reaches a peak and falls off at higher density. 
Therefore, the nature of total number of nucleons curve is mostly due to number of neutrons. Both the nucleon numbers go down for high temperature matter.  
At a particular density, the proton number goes up slightly for
$B_* = 10^{4}$ compared to non-magnetic case. This is the consequence
of higher values of proton fraction in the magnetic case as noted above.
It is also seen that the number of neutrons and as a result the number
of nucleons in the cell is smaller for $B_*=10^4$ at any given density.

\begin{figure}
    \centering
    \begin{tabular}{cc}
    \includegraphics[width=0.55\textwidth,angle=-90]{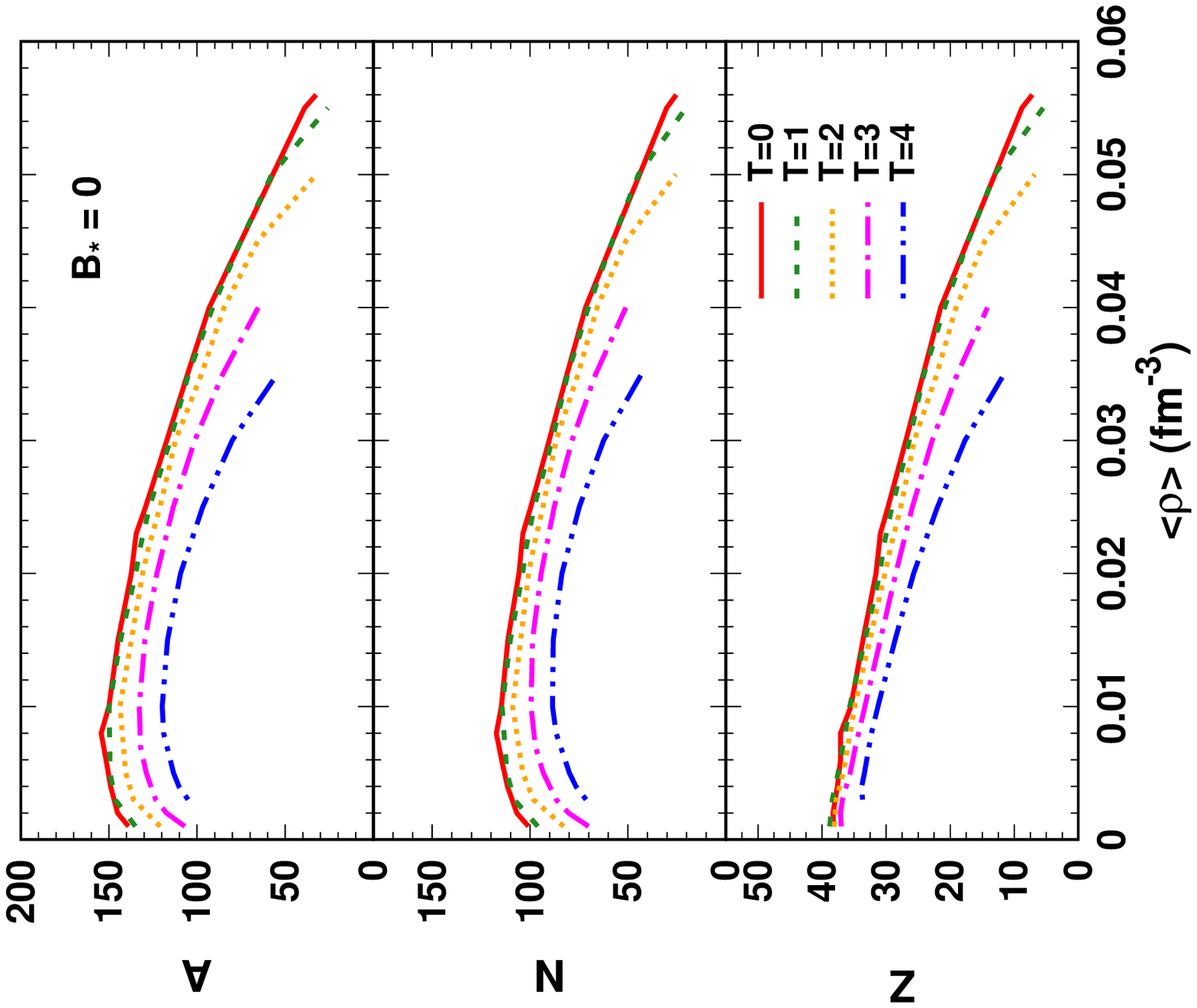} &
    \includegraphics[width=0.55\textwidth,angle=-90]{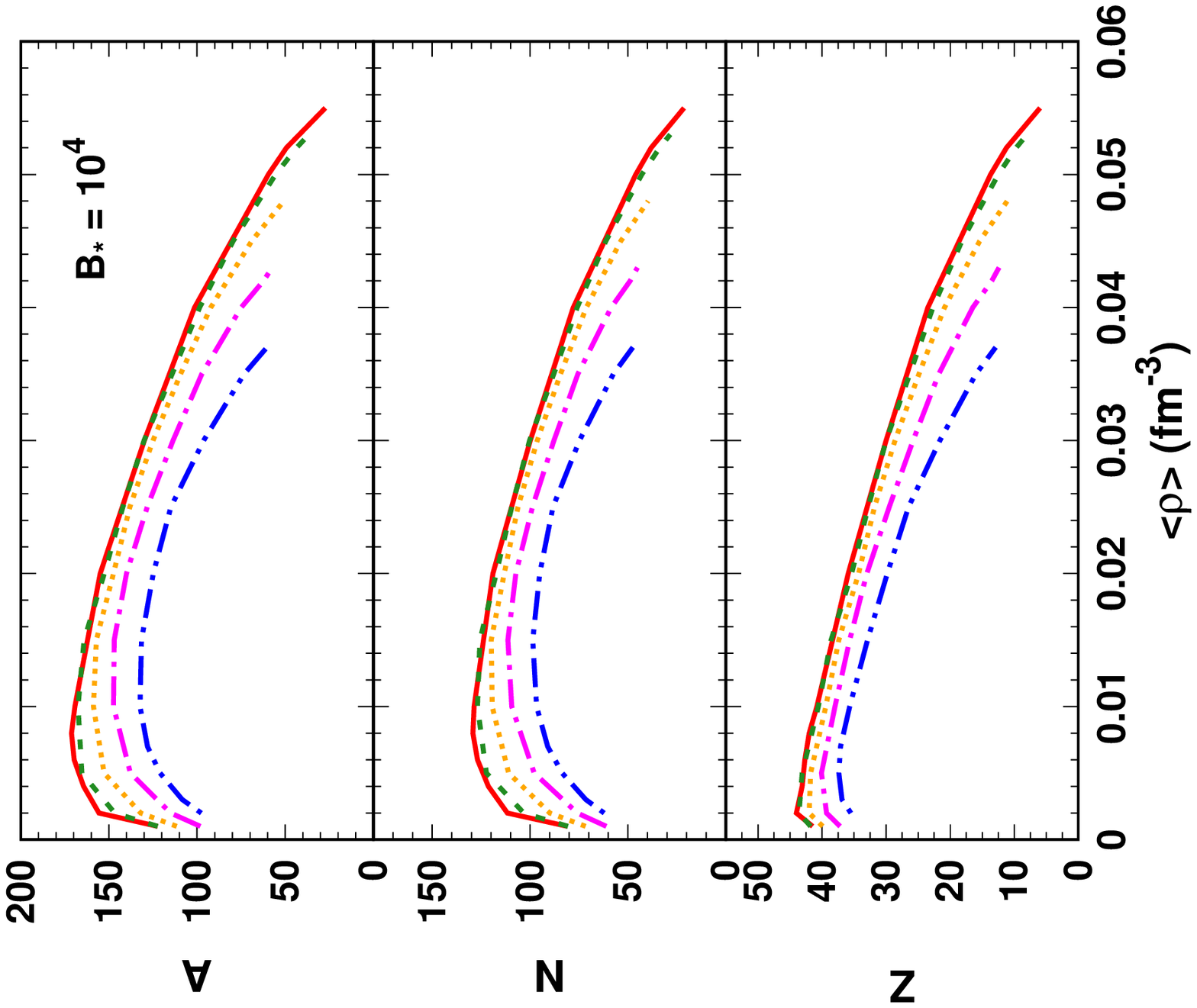}
    \end{tabular}
    \caption{Total number of neutrons ($N$), protons ($Z$) and $A=N+Z$ in the nucleus as a function of average baryon density and temperature for $B_*=0$ (left panel) and $B_*=10^4$
    (right panel).}
    \label{fig:AN}
\end{figure}
We plot the total number of nucleons, protons and neutrons inside the nucleus in Fig. \ref{fig:AN}  for non-magnetic and magnetic cases.
These numbers are obtained from the subtracted densities, using
Eq. \ref{eq:nz}.
Number of neutrons and as a consequence $A$ increase in the lower density part to reach a maximum, then they fall with increasing density for the entire range of temperature from $T=0$ to 4 MeV, for both the cases. The nucleus becomes smaller with temperature containing less numbers of neutrons and protons. These happen
because with rise in density and temperature increasing number
of neutrons drip out of the nucleus.
In presence of magnetic field of strength $B_*=10^4$ nuclei are
found to be heavier having larger number of neutrons and protons
at all densities and temperatures, as compared to the non-magnetic
case. 
This is again due to the extension of neutron drip point
and enhancement in proton fraction induced by the quantizing magnetic field.

\begin{figure}
    \centering
    \begin{tabular}{cc}
      \includegraphics[width=0.5\textwidth]{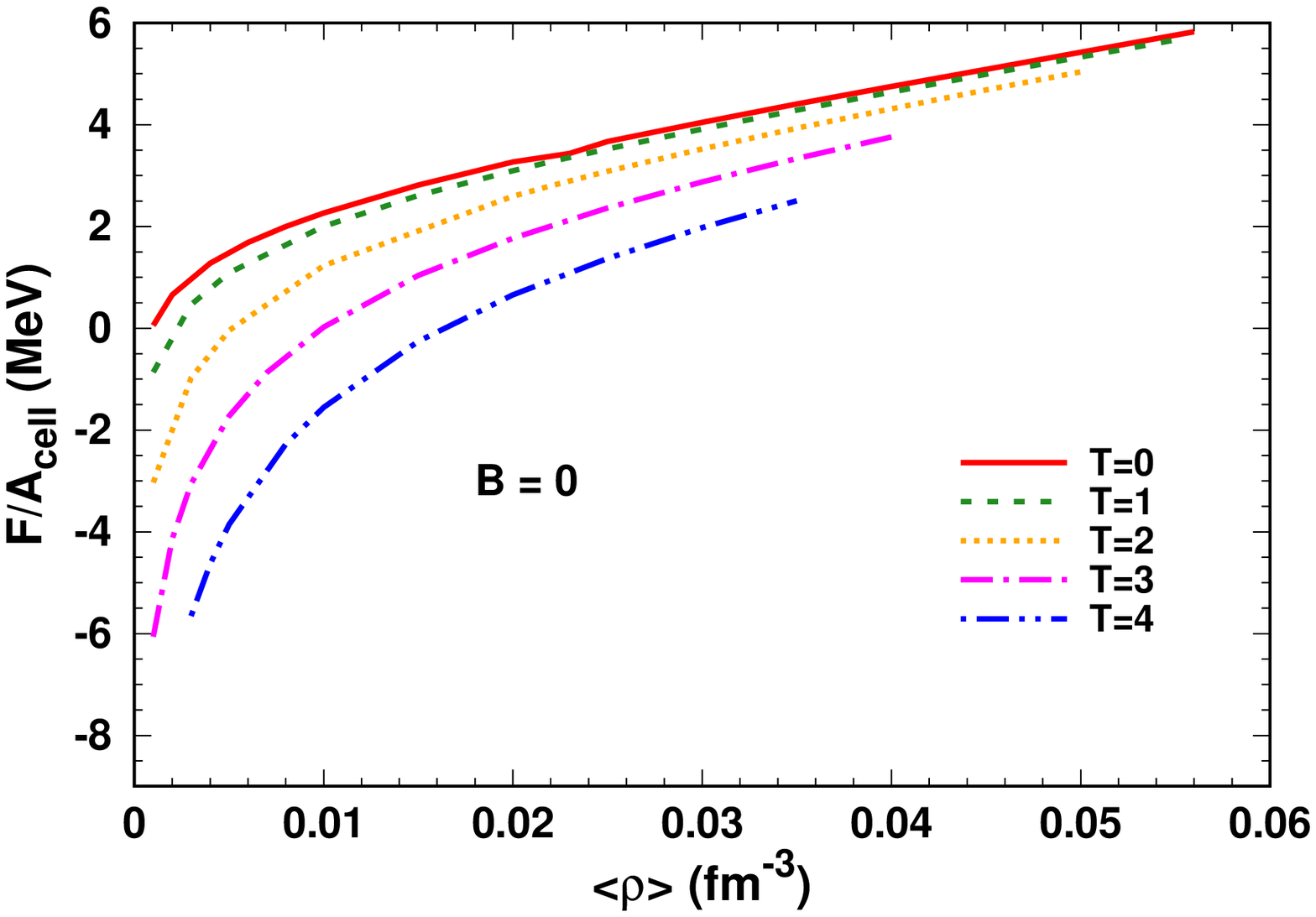} &
     \includegraphics[width=0.5\textwidth]{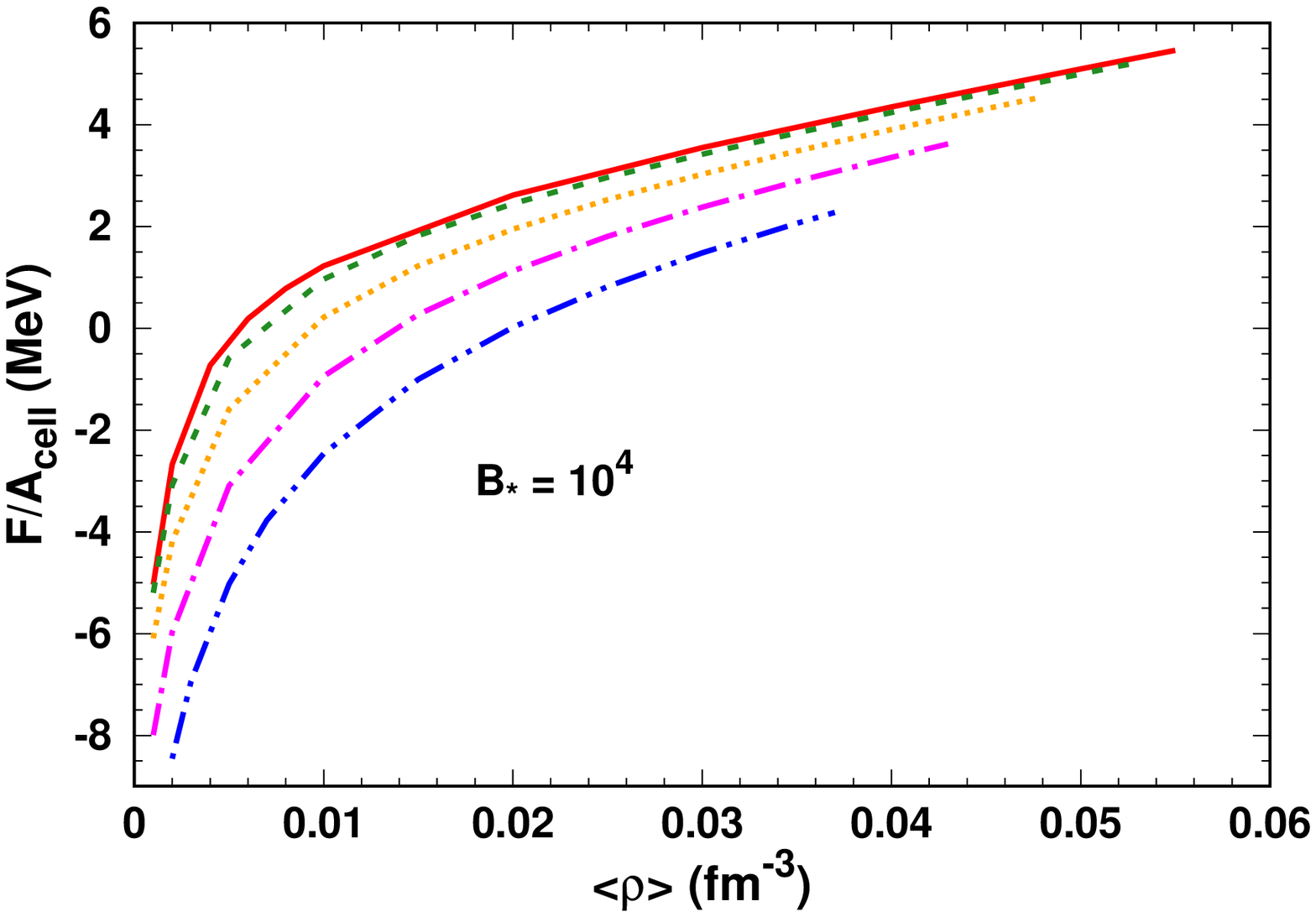} 
    \end{tabular}
    \caption{Free energy per baryon as function of baryonic number density for zero and finite temperature in presence and absence of quantizing magnetic field}
    \label{fig:Freeenergy}
\end{figure}

Finally, in Fig. \ref{fig:Freeenergy}, we plot the minimized free energy per nucleon ($F/A_{\rm cell}$) of the system with and without magnetic field, which increases monotonically with  baryon density.
On the other hand, at any given baryon density $F/A_{\rm cell}$ is maximum at $T=0$, and decreases as the temperature increases. 
This is mostly due to the last term in Eq. \ref{eq:facell}, which
grows with temperature.
For $B* = 10^4$  the $F/A_{\rm cell}$ values (see the right panel of Fig. \ref{fig:Freeenergy}) are smaller than the field free results 
both for $T=0$ and $T>0$ cases.
The lower $F/A_{\rm cell}$ values for the magnetized crust at the same density and temperature again emphasizes the greater binding energy of crustal matter in strong quantizing magnetic field \cite{Nandi2011}.

\section{Summary and Discussions}
We have studied the effect of finite temperature and strong magnetic
fields on the properties of neutron star inner crusts. 
We adopt the WS cell approximation where the nucleus
is considered to lie at the cell center and immersed in gases of free electrons and neutrons.  For the calculation of nuclear 
energy we use Skyrme energy density functional with SkM* interaction.
The equilibrium properties at a given density, temperature
and magnetic field are obtained by minimizing the free energy per nucleon of the cell under the condition of charge neutrality and
$\beta-$equilibrium. In order to isolate the properties of
equilibrium nucleus we employ the separation procedure within TF
formalism. Magnetic fields directly affects the electrons as their
motion perpendicular to the direction of the field get quantized in
Landau levels. The effect is significant for $B\gtrsim 10^{17}$ G, when 
electrons populate only the first Landau level. This results in less
number of dripped neutrons, higher proton fraction, heavier nucleus
and higher binding energy in the inner crust as compared to non-magnetic
case. However, the effect of temperature is found to act in the opposite
direction and reduces the impact of magnetic fields. We also find
that with increasing temperature the transition to uniform
matter takes place at lower density.

The dynamical ejecta of a binary neutron star merger can have two
components capable of synthesizing heavy elements  via
$r-$process \cite{Cowan:2019pkx}. 
One component is ejected because of the tidal forces and
contains very neutron-rich matter emanating from cold neutron star
crust. The other component is hotter as it originates due to the
shock heating at the interface of two merging neutron stars. If the
neutron stars possess strong magnetic fields or their initial low
fields get amplified during the merging process \cite{Ciolfi:2020cpf},
our results can be useful to calculate the $r-$process yields in both the
scenarios. 
However, at finite temperature single-nucleus description may not 
be adequate and one needs to consider mixtures of different
nuclei \cite{Lattimer:1985zf,Hempel:2009mc}. This is also manifested
in our calculation as we find that the free energy does not change
much with $A$ or $A_{\rm cell}$ (see Fig. \ref{fig:energies_B0}
and \ref{fig:energies_B10k}) when the temperature is high.
In a future study, we plan to investigate the effect of strong magnetic fields on the  composition of hot neutron star crust within a model that
would allow mixture of different nuclei.

\section{Acknowledgements}
\label{sec:ack}
SM and SB would like to acknowledge the  support of DAE-BRNS Research grant 37(3)/14/12/2018-BRNS. SM also thanks the hospitality of the Department of Physics, BITS Pilani, Hyderabad Campus where the research work was carried out. 
\\

\end{document}